\documentclass[conference]{IEEEtran}
\IEEEoverridecommandlockouts
\usepackage{cite}
\usepackage{amsmath,amssymb,amsfonts}
\usepackage{graphicx}
\usepackage{listings}
\usepackage{textcomp}
\usepackage{algorithm}
\usepackage{xcolor}
\usepackage{tcolorbox}
\usepackage{booktabs}
\usepackage{enumitem}
\usepackage{nicefrac}
\usepackage[hidelinks,bookmarks=false]{hyperref}
\usepackage{amsmath}
\usepackage{amsthm}
\usepackage{amsfonts}
\usepackage{amscd}
\usepackage{algpseudocode}
\usepackage{pgfplots}
\pgfplotsset{compat=1.8}
\usepgfplotslibrary{statistics}
\def\BibTeX{{\rm B\kern-.05em{\sc i\kern-.025em b}\kern-.08em
    T\kern-.1667em\lower.7ex\hbox{E}\kern-.125emX}}

\begin{document}
\begin{sloppy}

\title{Automatically Generating Documentation for Lambda Expressions in Java}

\author{\IEEEauthorblockN{Anwar Alqaimi\IEEEauthorrefmark{1},
Patanamon Thongtanunam\IEEEauthorrefmark{2} and
Christoph Treude\IEEEauthorrefmark{1}}
\IEEEauthorblockA{\IEEEauthorrefmark{1}School of Computer Science\\University of Adelaide\\
anwaribrahim.alqaimi@student.adelaide.edu.au, christoph.treude@adelaide.edu.au}
\IEEEauthorblockA{\IEEEauthorrefmark{2}School of Computing and Information Systems\\University of Melbourne\\
patanamon.thongtanunam@unimelb.edu.au}
}

\maketitle

\begin{abstract}

When lambda expressions were introduced to the Java programming language as part of the release of Java 8 in 2014, they were the language's first step into functional programming. Since lambda expressions are still relatively new, not all developers use or understand them. In this paper, we first present the results of an empirical study to determine how frequently developers of GitHub repositories make use of lambda expressions and how they are documented. We find that 11\% of Java GitHub repositories use lambda expressions, and that only 6\% of the lambda expressions are accompanied by source code comments. We then present a tool called \textsc{LambdaDoc} which can automatically detect lambda expressions in a Java repository and generate natural language documentation for them. Our evaluation of \textsc{LambdaDoc} with 23 professional developers shows that they perceive the generated documentation to be complete, concise, and expressive, while the majority of the documentation produced by our participants without tool support was inadequate. Our contribution builds an important step towards automatically generating documentation for functional programming constructs in an object-oriented language.

\end{abstract}


\section{Introduction and Motivation}
\label{sec:intro}

Modern programming languages enhance existing features and add new ones on a regular basis. For example, as part of the release of Java 8 in 2014, several new features and enhancements were introduced, ranging from improved type inference and method parameter reflection to the introduction of lambda expressions. According to Oracle,\footnote{\url{https://www.oracle.com/technetwork/java/javase/8-whats-new-2157071.html}, last accessed 5 Jan 2019.} lambda expressions ``enable you to treat functionality as a method argument, or code as data. Lambda expressions let you express instances of single-method interfaces (referred to as functional interfaces) more compactly.''

The introduction of lambda expressions to Java was motivated by the suitability of functional programming for synchronised, parallel, and event-driven programming. Other programming languages, including Groovy, Scala, and Python, already supported functional programming. Lambda expressions are useful since they are shorter than anonymous classes and can result in a similar outcome. 

The word lambda is derived from the Greek letter lambda ($\lambda$) to represent a function of the abstract~\cite{Sharan2014}. Lambda expressions in Java can be defined as nameless code blocks of functions that contain a collection of formal parameters and a body that is joined by an arrow (\texttt{->}).
Listing \ref{lst:lambda_example} shows examples of lambda expressions in Java~\cite{Sharan2014}.
A lambda expression that declares its parameters types is called an explicit lambda expression and a lambda expression that does not declare its parameters types is called an implicit lambda expression~\cite{Sharan2014}. 

\begin{lstlisting}[language=java, caption=Examples of lambda expressions, label=lst:lambda_example]
//Ex1: Explicit lambda expression
//Takes an integer parameter and returns 
//the parameter value incremented by 1
(int x) -> x + 1

//Ex2: Implicit lambda expression
//Takes two integer parameters and returns
//the maximum of the two
(x, y) -> 
{int max = x > y ? x : y;
return max;}
\end{lstlisting}

One of the primary objectives of using lambda expressions is to maintain a short syntax and to allow the compiler to deduce the details. However, along with the compiler, developers are also often left to deduce the details on their own. Understanding lambda expressions is not trivial, and good software documentation has long been known to be rare~\cite{Lethbridge2003}. To make lambda expressions more accessible to developers, we follow in a long line of work on automatically documenting particular parts of source code (e.g., test cases~\cite{Li2016} and database code~\cite{Linares-Vasquez2016}) to develop a novel approach called \textsc{LambdaDoc} to automatically document lambda expressions in Java. To the best of our knowledge, \textsc{LambdaDoc} is the first tool of \mbox{its nature}.

In this paper, we contribute:

\begin{itemize}
\item An empirical study of how lambda expressions are used in 435 engineered software repositories hosted on GitHub.
\item An empirical study of the documentation of lambda expressions in GitHub repositories via source \mbox{code comments}.
\item \textsc{LambdaDoc}, a novel approach to automatically document lambda expressions in Java, along with an evaluation with 23 professional developers.
\end{itemize}

The remainder of this paper is structured as follows. We introduce our research questions as well as our methods for data collection and analysis in Section~\ref{sec:method}. \textsc{LambdaDoc} is described in detail in Section~\ref{sec:lambdadoc} and our findings are reported in Section~\ref{sec:findings}, separately for each of our research questions. We discuss threats to validity in Section~\ref{sec:threats} and related work in Section~\ref{sec:related}, before we conclude the paper and outline future work in Section~\ref{sec:conclusions}.

\section{Research Method}
\label{sec:method}

In this section, we outline our research questions as well as the data collection and analysis methods used to answer them.

\subsection{Research Questions}
\label{sec:rqs}

To study how lambda expressions are used by Java developers, we first needed to devise an approach for detecting lambda expressions in non-compilable code and code snippets in commits. We chose not to rely on existing static analysis approaches (e.g., PMD~\cite{Louridas2006}) since these approaches often require compilable code to identify lambda expressions.
To evaluate our custom-built lambda expression detector, we first addressed the following research question:

\begin{description}
\item{\textbf{RQ1}} How accurate is our approach for detecting lambda expressions?
\end{description}

Since lambda expressions have been recently introduced in Java, little is known about the prevalence of lambda expression usage in Java projects. Hence, we aimed to establish how frequently lambda expressions are used by Java developers. In addition, we investigated the kind of lambda expressions that are commonly used:

\begin{description}
\item{\textbf{RQ2}} How are lambda expressions used?
\begin{description}
\item{\textbf{RQ2.1}} How many repositories use lambda expressions, and what is the amount of lambda expressions per repository?
\item{\textbf{RQ2.2}} Are the same lambda expressions used in multiple repositories?
\end{description}
\end{description}

In addition to establishing how Java developers are making use of lambda expressions, a goal of our work was the automated generation of documentation for these expressions. To inspire our work on documentation generation, we next asked how lambda expressions are currently documented in the form of source code comments: 

\begin{description}
\item{\textbf{RQ3}} How are lambda expressions documented?
\begin{description}
\item{\textbf{RQ3.1}} How many lambda expressions are accompanied by comments?
\item{\textbf{RQ3.2}} What kind of comments accompany lambda expressions?
\end{description}
\end{description}

Finally, to evaluate our documentation generation approach for lambda expressions \textsc{LambdaDoc}, we asked professional developers to evaluate the automatically \mbox{generated documentation}:

\begin{description}
\item{\textbf{RQ4}} How well can \textsc{LambdaDoc} document lambda expressions?
\begin{description}
\item{\textbf{RQ4.1}} How do developers document lambda expressions when asked to provide comments?
\item{\textbf{RQ4.2}} How is the documentation generated using \textsc{LambdaDoc} perceived by developers?
\end{description}
\end{description}

\subsection{Data Collection}
\label{sec:collection}

\subsubsection{Repositories}

Motivated by previous work on the perils of mining GitHub~\cite{Kalliamvakou2014}, e.g., a large portion of repositories on GitHub are not for software development, we used the RepoReaper framework developed by Munaiah et al.~\cite{Munaiah2017} to select repositories for our work. RepoReaper was developed to address the difficulty to differentiate between repositories with engineered software projects and those with assignments and noise. The ratio of unwanted repositories in a stochastic sample could distort research and cause illogical and possibly incorrect conclusions. RepoReaper contains repositories classified as organisation and utility. 

To select repositories for our study, we first obtained all 51,392 Java repositories which had been classified as containing engineered software projects by the Random Forest classification of RepoReaper.
We then randomly sampled from these 51,392 repositories in batches of 1,000 until we had obtained at least 400 repositories which contained at least one lambda expression detected by our lambda expression detector (see Section \ref{sec:detect_lambda}). This way, we were able to ensure that our conclusions concerning the ratio of repositories with a specific characteristic would generalise to the entire population of engineered Java repositories on GitHub containing lambda expressions with a confidence level of 95\% and a confidence interval of 5.\footnote{\url{https://www.surveysystem.com/sscalc.htm}, last accessed 5 Jan 2019.} After cloning and analysing 4,000 repositories (i.e., four batches of 1,000, the number it took to find at least 400 repositories containing at least one lambda expression), we had retrieved a total of 435 repositories containing lambda expressions, i.e., 11\%. These 435 repositories are a statistically representative sample of all engineered Java repositories containing lambda expressions. They contained a total of 497,108 Java files, out of which 9,933 contained at least one lambda expression.
In total, we collected 54,071 lambda expressions across the 435 Java repositories.

\subsubsection{Detection of Lambda Expressions}
\label{sec:detect_lambda}
We developed an approach to detect lambda expressions and collect their metadata, i.e., start line number, start character position, end line number, end character position, number of lines, number of parameters, and type (explicit vs.~implicit).

To identify lambda expressions in the source code of each Java file, our lambda expression detector first reads source code line-by-line until encountering a lambda arrow (\texttt{->}) that is not part of a comment or a string. 
The detector then checks whether the lambda expression covers multiple lines---a multi-line lambda expression starts with an open-parenthesis or an open-bracket and ends with the corresponding closing symbol while a single-line lambda expression ends with a semicolon or a parenthesis. The detector then determines if the lambda expression is explicit or implicit by checking whether the parameter list contains parameter types. 
For example, in Listing~\ref{lst:lambda_example},  the lambda expression \textit{Ex1} is identified as a single-line lambda expression that has an implicit type, while the lambda expression \textit{Ex2} is identified as a multi-line lambda expression that has an explicit type.
Finally, our lambda expression detector extracts source code comments written directly above the lambda expression.
For multi-line lambda expressions, our detector also extracts source code comments written within the lambda expression.

\subsubsection{Practitioner Survey}

To evaluate the documentation generated by our automated approach, we employed a practitioner survey which followed a similar structure used by Linares-V\'{a}squez et al.~\cite{Linares-Vasquez2016}.
We recruited participants through Amazon Mechanical Turk\footnote{\url{https://www.mturk.com/}, last accessed 22 Jan 2019.} and the required qualification was ``Employment Industry -- Software \& IT Services''.
This methodology has been successfully used by previous work~\cite{Prana2019}. 


\begin{table*}
\centering
\caption{Survey questions. Each horizontal like indicates a page break of the survey. Note that Questions 9--12 were repeated five times for each of the lambda expressions and corresponding documentation shown in Table~\ref{tab:fivesummaries}.}
\label{tab:surveyquestions}
\begin{tabular}{lp{8.25cm}p{8.25cm}}
\toprule
 & Question & Answer options \\
\midrule
1 & Is developing software part of your job? & Yes / No \\
2 & What is your job title? & Open-ended \\
3 & For how many years have you been developing software? & Less than one year / 1--2 years / 2--4 years / 4--6 years / More than 6 years \\
4 & How many years of experience do you have in Java development? & Less than one year / 1--2 years / 2--4 years / 4--6 years / More than 6 years \\
5 & How would you rate your expertise in Java? & Beginner / Intermediate / Expert \\
6 & How confident are you in your ability to WRITE lambda expressions in Java? & Not confident at all / Slightly confident / Somewhat confident / Fairly confident / Completely confident \\
7 & How confident are you in your ability to READ lambda expressions in Java? & Not confident at all / Slightly confident / Somewhat confident / Fairly confident / Completely confident \\
8 & What is your area of software development? (e.g., web, systems, embedded) & Open-ended \\
\midrule
9 & Consider the following lambda expression in Java. Could you please write a one-sentence summary of what this lambda expression does? & Open-ended \\
\midrule
10 & Consider now the following sentence which aims to describe the lambda expression. Only focusing on the content of the sentence without considering the way it has been presented, do you think the description is COMPLETE? & The sentence is considered to be complete / The sentence misses some important information to understand the lambda expression / The sentence misses the majority of the important information to understand the lambda expression \\
11 & Consider the same lambda expression and sentence. Only focusing on the content of the sentence without considering the way it has been presented, do you think the sentence is CONCISE? & The sentence is considered to be concise / The sentence contains some redundant/useless information / The sentence contains a lot of redundant/useless information \\
12 & Consider the same lambda expression and sentence. Only focusing on the content of the sentence without considering the completeness and conciseness, do you think the sentence is EXPRESSIVE? & The sentence is easy to read and understand / The sentence is somewhat readable and understandable / The sentence is hard to read and understand \\
\midrule
13 & Please try \textsc{LambdaDoc} for any lambda expressions of your choice at $<$link to \textsc{LambdaDoc} web application$>$. Please provide feedback on \textsc{LambdaDoc} here. & Open-ended \\
14 & How likely is it that you would recommend this tool to a friend or colleague? & 5-point Likert scale from ``not likely at all'' to ``extremely likely'' \\
15 & How would you prefer this tool to be implemented? & As an Eclipse plugin / As a GUI Application / As a website/web service / Other (multiple-choice) \\
16 & Which of the following software engineering tasks would you use this type of documentation of lambda expressions in Java for? & Implementation / Testing / Documentation / Maintenance / Other (multiple-choice) \\
17 & Do you have any further comments about lambda expressions or this survey? & Open-ended \\
\bottomrule
\end{tabular}
\end{table*}

\begin{table*}
\centering
\caption{Lambda expressions and corresponding documentation generated by \textsc{LambdaDoc}}
\label{tab:fivesummaries}
\begin{tabular}{p{8.65cm}p{8.65cm}}
\toprule
Lambda expression & Generated documentation \\
\midrule
\vspace{-1.5\baselineskip} \begin{lstlisting}[language=java] 
callInContext( REPO_USER_2, repo2.getId(), MASTER_BRANCH, () -> createNode ( NodePath.ROOT, "repo2Node" ) );
\end{lstlisting} & This lambda expression does not take any parameter and returns the result of the execution of the ``create Node'' method with two parameters ``NodePath ROOT and ``repo2Node''~''. \\
\vspace{-1.5\baselineskip} \begin{lstlisting}[language=java]
(Integer t, Integer t1) -> Double.compare(splitEvaluation[t], splitEvaluation[t1]) \end{lstlisting} & This lambda expression takes 2 parameters Integer t and Integer t1 and returns the result of the execution of Double's ``compared to'' method with two parameters element of ``split Evaluation'' array t and element of ``split Evaluation'' array t1. \\
\vspace{-1.5\baselineskip} \begin{lstlisting}[language=java]
.peek(batch->count3= count3+batch.size()) \end{lstlisting} & This lambda expression takes 1 parameter batch and returns count3 equal count3 plus the result of the execution of the ``size'' method on it. \\
\vspace{-1.5\baselineskip} \begin{lstlisting}[language=java]
.beforeResolved(ExecutableComponent.class, ec -> ec.set("c")) \end{lstlisting} & This lambda expression takes 1 parameter ec and returns the result of the execution of the ``set'' method on it with parameter ``c''. \\
\vspace{-1.5\baselineskip} \begin{lstlisting}[language=java]
return stream.flatMap(t -> Stream.of(value, t)) \end{lstlisting} & This lambda expression takes 1 parameter t and returns the result of the execution of Stream's ``of'' method with two parameters ``value and t''. \\
\bottomrule
\end{tabular}
\end{table*}

Table~\ref{tab:surveyquestions} shows our survey questions as well as the answer options for each one. 
The first section of the survey (Questions 1--8) collected demographic information and established the participants' experience with software development, Java, and reading and writing lambda expressions. 
Then, Questions 9--12 were used to evaluate our generated documentation for each of the five lambda expressions that were randomly sampled from all detected lambda expressions of this work (see Table~\ref{tab:fivesummaries}).
Question 9 asked participants to write a summary of what the lambda expression shown in the question does. 
Then, Questions 10--12 asked participants to evaluate our generated documentation for the lambda expression shown in Question 9.
The generated documentation was evaluated in terms of completeness, conciseness, and expressiveness. 
Note that participants were not able to see our generated documentation until they had answered Question 9.
Table~\ref{tab:fivesummaries} shows the lambda expressions and our generated documentations that we used for our evaluation.
The last section of the survey provided participants with a link to a web application of \textsc{LambdaDoc} where users could submit a lambda expression and have documentation generated for it.  
Finally, Questions 13--17 asked participants about the usefulness of our tool.


\begin{figure}
\centering
\begin{tikzpicture}
\begin{axis}[
    y = 0.2cm,
    ylabel = participants,
    xlabel = software development experience in years,
    symbolic x coords={0--1, 1--2, 2--4, 4--6, 6+},
    xtick=data,
    ymin = 0]
    \addplot[ybar,fill=white] coordinates {
    (0--1, 2)
    (1--2, 5)
    (2--4, 8)
    (4--6, 4)
    (6+, 4)
    };
\end{axis}
\end{tikzpicture}
\caption{Software development experience of survey participants}
\label{fig:softwaredevexperience}
\end{figure}
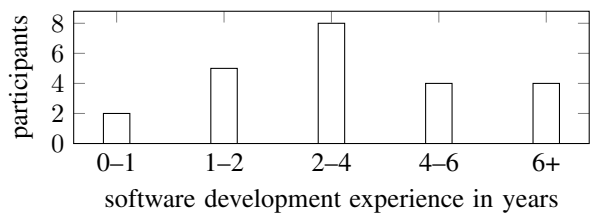

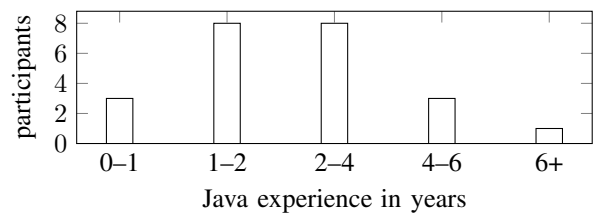
\begin{figure}
\centering
\begin{tikzpicture}
\begin{axis}[
    y = 0.2cm,
    ylabel = participants,
    xlabel = Java experience in years,
    symbolic x coords={0--1, 1--2, 2--4, 4--6, 6+},
    xtick=data,
    ymin = 0]
    \addplot[ybar,fill=white] coordinates {
    (0--1, 3)
    (1--2, 8)
    (2--4, 8)
    (4--6, 3)
    (6+, 1)
    };
\end{axis}
\end{tikzpicture}
\caption{Java experience of survey participants}
\label{fig:javaexperience}
\end{figure}

We obtained responses from 23 participants. Only one participant indicated that developing software was not part of their job. The job titles of participants varied from ``Data administrator'' to ``Junior software engineer'', but all titles were related to software development or information technology. Most participants specified either web development (\nicefrac{8}{23} = 35\%) or backend/systems development (\nicefrac{7}{23} = 30\%) as their area of software development, with other answers indicating various areas from computer vision to databases. The majority of participants had considerable software development and Java experience, as shown in Figures~\ref{fig:softwaredevexperience} and~\ref{fig:javaexperience}, i.e., 70\% and 52\% of participants had software development and Java experience of at least 2 years, respectively. Participants considered their expertise in Java to either be at the level of beginners (\nicefrac{10}{23} = 43\%) or intermediate (\nicefrac{11}{23} = 48\%), with only 2 experts (9\%).
Table~\ref{tab:confidence} shows that a considerable group of participants considered themselves fairly confident in both reading and writing lambda expressions (\nicefrac{8}{23} = 35\%), with other answers varying between slightly and somewhat confident. None of the participants indicated to be completely confident in their abilities to either read or write lambda expressions, further motivating our work. Only 2 participants (9\%) considered themselves not confident at all. 

\begin{table}
\centering
\caption{Participant confidence in reading and writing lambda expressions}
\label{tab:confidence}
\begin{tabular}{llr}
\toprule
Confidence in reading & Confidence in writing & Count \\
\midrule
Not confident at all & Not confident at all & 2 \\
Slightly confident & Slightly confident & 2 \\
Fairly confident & Slightly confident & 2 \\
Slightly confident & Somewhat confident & 4 \\
Somewhat confident & Somewhat confident & 3 \\
Somewhat confident & Fairly confident & 2 \\
Fairly confident & Fairly confident & 8 \\
\bottomrule
\end{tabular}
\end{table}

\subsection{Data Analysis}
\label{sec:analysis}

In this section, we outline the data analysis methods used to answer our research questions.

\subsubsection{Accuracy of lambda expression detection}

To evaluate the accuracy of our tooling to detect lambda expressions in source code files and commits (i.e., lambda expression detector), two authors of this paper who were not involved in the implementation of the lambda expression detector manually annotated a sample of lines from source code files and commits which contained an arrow (\texttt{->}) to indicate whether the arrow marked the beginning of a lambda expression or served some other function (e.g., as part of a source comment or string). We hypothesise that detecting lambda expressions in commits is harder than in source code files since a developer might only commit part of a lambda expression, e.g., for those expressions which span multiple lines.

The two authors independently annotated 100 such randomly sampled lines from source code files and 100 such randomly sampled lines from commits. They achieved perfect agreement (100\%), and one of them further annotated 300 lines from each set. Based on the annotation of a total of 400 lines from source code files and 400 lines from commits, our conclusions regarding the accuracy of the lambda expression detection generalise to the population of all lines containing arrows with a confidence interval of 5 at a confidence level of 95\%.\footnote{\url{https://www.surveysystem.com/sscalc.htm}, last accessed 5 Jan 2019.} We then compared the manual annotation with the results from our lambda expression detector.

\subsubsection{Lambda expression usage}

To analyse the frequency with which lambda expressions are used in GitHub repositories, we analysed descriptive statistics of our data set. To investigate whether there are lambda expressions which are used in multiple repositories, we examined the number of unique lambda expressions.
Note that lambda expressions were normalised by removing white space before analysing unique lambda expressions and that we compared lambda expressions textually.
Furthermore, we investigated the characteristics of lambda expressions by analysing their metadata, e.g., the number of parameters and types.


\subsubsection{Documentation of lambda expressions}

To investigate the extent to which lambda expressions are already documented in GitHub repositories, we quantitatively analysed how many of the lambda expressions were accompanied by a comment either directly above the expression (i.e., \textit{above comments}) or within the expression (i.e., \textit{within comments}). We then qualitatively analysed a statistically representative sample of the comments we found in order to determine whether the comments actually described the functionality of the lambda expressions. For the qualitative annotation, one of the authors established a coding scheme based on a preliminary analysis of 100 comments. Another author then used this coding schema on the same data, allowing us to calculate inter-rater agreement. Two authors then applied this coding schema to annotate a total of 200 above comments and 200 within comments.

\subsubsection{\textsc{LambdaDoc} evaluation}

To explore how developers document lambda expressions, we qualitatively analysed the responses to Question 9 of our survey (cf.~Table~\ref{tab:surveyquestions}). Since the goal of this analysis was to establish the level of detail with which developers document lambda expressions, we used the pre-defined categories ``adequate'', ``incomplete'', and ``inadequate'' for the annotation. One author annotated all 115 documentation attempts (23 participants $\times$ 5 \mbox{documentation attempts}). 

Finally, to understand the perceptions of developers regarding our tool \textsc{LambdaDoc}, we quantitatively analysed the responses to survey Questions 10--16 (cf.~Table~\ref{tab:surveyquestions}).

\section{\textsc{LambdaDoc}}
\label{sec:lambdadoc}

\begin{algorithm}
	\caption{\textsc{LambdaDoc}(lambdaExpression)}
	\label{alg:lambdadoc}
	\begin{algorithmic}[1]
		\State $\text{doc} \gets \emptyset$
		\State $\text{intro} \gets \text{``This lambda expression''}$
		\State $\text{exp} \gets \text{lambdaExpression.split(``$->$'')}$
		\State $\text{pc} \gets \text{numberOfParameters(exp[0])}$
		\If {$\text{pc} == 0$}
		\State $\text{pText} \gets \text{``does not take any parameter''}$
		\Else
		\If {$\text{pc} == 1$}
		\State $\text{p} \gets \text{parameterName(exp[0])}$
		\State $\text{pText} \gets \text{``takes 1 parameter'' + p}$
		\Else	
		\If {$\text{pc} > 1 $}
		\State $\text{ps} \gets \text{parameterNames(exp[0])}$
		\State $\text{pText} \gets \text{``takes'' + pc + ``parameters'' + ps}$
		\EndIf 
		\EndIf 
		\EndIf
		\If {$\text{exp[1].contains(``.'')}$}
		\State $\text{i} \gets \text{0}$
		\State $\text{mText} \gets \emptyset$
		\For{$\text{i} < \text{exp[1].length()}$} 
		\If {$\text{exp[1].charAt(i) == operator}$}
		\State $\text{mText} \gets \text{mText + operatorToWord(operator)}$
			\State $\text{i}++$
		\EndIf
		\If {$\text{exp[1].charAt(i)} == \text{``.''} $}
		\State $\text{dot} \gets \text{i}$
		\While{$\text{exp[1].charAt(i)} != \text{``(''}$}
		\State $\text{mText} \gets \text{mText + exp[1].charAt(i)}$
			\State $\text{i}++$
		\EndWhile
		\State $\text{j} \gets \text{dot}$
		\State $\text{oText} \gets \emptyset$
		\For{$\text{j} > 0$} 
		\While{$\text{exp[1].charAt(j).isDigitOrNum}$}
		\State $\text{oText} \gets \text{oText + exp[1].charAt(j)}$
		\State $\text{j}--$
		\EndWhile
		\State $\text{oText} \gets \text{reverseString(oText)}$
		\State $\text{mText} \gets \text{camelCaseSplit(mText)}$
		\State $\text{rText} \gets \text{oText + ``'s'' + mText)}$
		\EndFor
		\EndIf
		\EndFor
		\State $\text{doc} \gets \text{intro + pText + ``and returns'' + rText}$
		\Else
		\While{$\text{exp[1].contains(operator)}$}
		\State $\text{rText} \gets \text{operatorToWord(rText)}$
		\EndWhile
		\State $\text{doc} \gets \text{intro + pText + ``and returns'' + rText}$
		\EndIf
		\State $\text{return doc}$
	\end{algorithmic}
\end{algorithm}

In this section, we present our \textsc{LambdaDoc} approach to generate documentation for a lambda expression. 
The documentation generated by \textsc{LambdaDoc} starts with the phrase ``This lambda expression'', followed by details of parameters and body of the lambda expressions. 
Table~\ref{tab:fivesummaries} shows examples of the documentation generated by \textsc{LambdaDoc}.

Algorithm~\ref{alg:lambdadoc} shows the algorithm of \textsc{LambdaDoc} in pseudocode.
The first step of the algorithm is to identify the parameters and the body of a lambda expression by dividing the expression at the arrow (\texttt{->}) and to determine the number of parameters.
If there are no parameters, the text ``does not take any parameter'' is added to the documentation.
Otherwise, the names of the parameters are determined, and a sentence fragment is generated which indicates the number of parameters along with their names.

The body is analysed to determine if it contains at least one method call based on the presence of a dot (\texttt{.}) not followed by a digit. If the body contains a method, its object, name, and parameters are identified. The method name is split based on camel case.
Then object, method name, and parameters are concatenated in the documentation. Operators (e.g., \texttt{+}) are replaced with a word representing them (e.g., ``plus'').

Note that \textsc{LambdaDoc} is designed to generate documentation for lambda expressions which contain a single statement.

\section{Findings}
\label{sec:findings}

In this section, we report the findings for our \mbox{research questions}.

\subsection{Accuracy of lambda expression detection}
\label{sec:accuracy}

Out of the 400 lines from source code files which contained an arrow (\texttt{->}), our manual annotation revealed that 258 (65\%) of the lines contained the start of a lambda expression while the remaining 142 (35\%) lines contained arrows for other reasons, mostly as part of strings or in source code comments.
Using the same set of source code lines, we used our lambda expression detector to identify whether the lines contained a lambda expression or not.
We found that our lambda expression detector achieved a recall of 1, i.e., all of the 258 lines containing lambda expressions from the manual annotation result were identified as containing lambda expressions by the lambda expression detector.
Furthermore, our lambda expression detector achieved a precision of 1, i.e., all of the lines that were identified as containing lambda expressions by our lambda expression detector are those 258 lines containing lambda expressions from the \mbox{manual annotation}. 

Similarly, our manual annotation of the 400 lines from commits which contained an arrow identified 269 (67\%) lines in which the arrow belonged to a lambda expression and 131 (33\%) lines where the arrow was not part of a lambda expression.
Based on this manual annotation result, our lambda expression detector achieved a recall and precision of 1. Note that this result could be impacted by lambda expressions which do not contain subsequent characters indicating arrows (\texttt{->}); however, we are not aware of such expressions.


\begin{tcolorbox}
\textbf{Summary:} Our approach for detecting lambda expressions is able to identify lambda expressions in source code files and commits with perfect precision and recall.
\end{tcolorbox}

\subsection{Lambda expression use}
\label{sec:frequency}

As reported in Section~\ref{sec:collection}, 11\% (\nicefrac{435}{4,000}) of the Java repositories in our sample made use of lambda expressions at least once.
For these 435 repositories, Figure~\ref{fig:lambdasperrepository} shows the number of lambda expressions that are contained in each repository. The distribution has a long tail---a few the repositories extensively use lambda expressions (e.g., 10\% of the repositories contain more than 100 expressions) while the majority of the repositories sometimes uses them. In our sample, aol/simple-react\footnote{\url{https://github.com/aol/cyclops}, last accessed 19 Jan 2019.} and elastic/elasticsearch\footnote{\url{https://github.com/elastic/elasticsearch}, last accessed 19 Jan 2019.} are the most prolific users of lambda expressions, with 18,754 and 11,886 expressions, respectively.

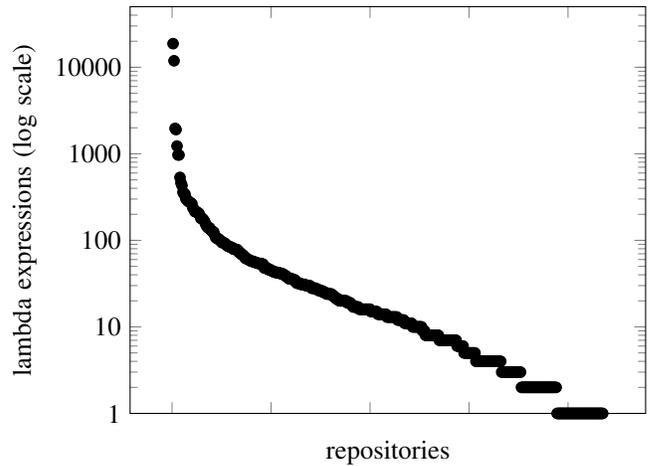
\begin{figure}
\centering
\begin{tikzpicture}
\begin{axis}[
y = 0.5cm,
yticklabels = {1, 10, 100, 1000, 10000},
ylabel = lambda expressions (log scale),
xlabel = repositories,
ymode = log,
xticklabels={,,},
ymin = 1]
\addplot[only marks] table [x index=0,y index=1,col sep=comma] {lambdasperrepository.dat};
\end{axis}
\end{tikzpicture}
\caption{Number of lambda expressions per repository, considering only the subset of repositories which use lambda expressions (log scale)}
\label{fig:lambdasperrepository}
\end{figure}

\begin{figure}
\centering
\begin{tikzpicture}
\begin{axis}[
y = 0.5cm,
yticklabels = {1, 10, 100},
ylabel = lines (log scale),
xlabel = lambda expressions,
ymode = log,
xticklabels={,,},
ymin = 1]
\addplot[only marks] table [x index=0,y index=1,col sep=comma] {linesperlambda.dat};
\end{axis}
\end{tikzpicture}
\caption{Number of lines per lambda expression (log scale)}
\label{fig:linesperlambda}
\end{figure}
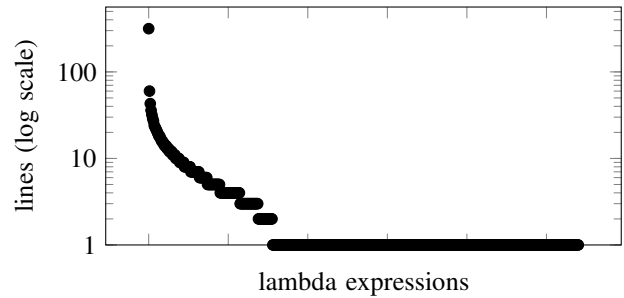

\begin{figure}
\centering
\begin{tikzpicture}
\begin{axis}[
y = 0.25cm,
ylabel = parameters,
xlabel = lambda expressions,
xticklabels={,,},
ymin = 0]
\addplot[only marks] table [x index=0,y index=1,col sep=comma] {paramsperlambda.dat};
\end{axis}
\end{tikzpicture}
\caption{Number of parameters per lambda expression}
\label{fig:paramsperlambda}
\end{figure}
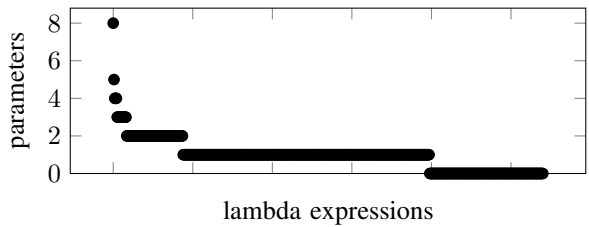

Out of the 54,071 lambda expressions in our data set, 33,916 were unique (after removing white space). Table~\ref{tab:commonlambdas} shows the three lambda expressions which were used in the largest number of the studied Java repositories.
As shown in the table, lambda expressions can play an important role in exception handling. An additional analysis of the 33,916 unique lambda expressions showed that 4,920 (15\%) contain the string ``exception''.

We now report common characteristics of lambda expressions in our data set.
Figure~\ref{fig:linesperlambda} shows that 70\% of the lambda expressions are single-line, while there are outliers with up to more than 300 lines per one lambda expression.
Figure~\ref{fig:paramsperlambda} shows that the majority of the lambda expressions (57\%) have exactly one parameter
while a sizeable minority of lambda expressions (27\%) does not have any parameters. 
The other lambda expressions have two to eight parameters.
We also found that 98\% of the lambda expressions were implicit, i.e., they do not declare their parameter types.

\begin{table}
\centering
\caption{Lambda expressions used across repositories}
\label{tab:commonlambdas}
\begin{tabular}{p{6.6cm}r}
\toprule
Lambda expression & Repositories \\
\midrule
\vspace{-1.5\baselineskip} \begin{lstlisting}[language=java] 
return () -> { try { return task.call(); } catch (Exception e) { handle(e); throw e; } };\end{lstlisting} & 7 \\
\vspace{-1.5\baselineskip} \begin{lstlisting}[language=java]
return () -> { try { task.run(); } catch (Exception e) { handle(e); } }; \end{lstlisting} & 7 \\
\vspace{-1.5\baselineskip} \begin{lstlisting}[language=java]
.map(user -> new ResponseEntity<>(user, HttpStatus.OK)) \end{lstlisting} & 7 \\
\bottomrule
\end{tabular}
\end{table}

\begin{tcolorbox}
\textbf{Summary:} 11\% of Java GitHub repositories use lambda expressions at least once. Exception handling is a common purpose of using lambda expressions. Lambda expressions are usually implicit, single-line, and have one parameter. 
\end{tcolorbox}

\subsection{Documentation of lambda expressions}
\label{sec:documentation}

Out of the 54,071 lambda expressions in our data set, the vast majority (50,984 = 94\%) was not accompanied by any documentation, neither right above the expression nor within the expression. We found 1,531  (3\%) lambda expressions with a comment right above, 1,298 (2\%) lambda expressions with a comment within, and an additional 258 (0.5\%) lambda expressions with a comment above and a comment within.

To understand whether the comments which accompany lambda expressions actually document the expression, we manually annotated a randomly sampled set of 200 comments from above a lambda expression and another 200 comments from within a lambda expression. One author annotated 50 comments from each set to establish the following \mbox{coding schema}:

\begin{itemize}
\item high-level documentation: the comment describes the lambda expression, but at a very high level. An example is the comment ``\texttt{// start bottom-up}'' above a 53-line lambda expression---while it captures the core purpose of the expression, it does not explain how this purpose was achieved.
\item reasonably detailed explanation: the comment appears to explain the lambda expression reasonably well. An example is the comment ``\texttt{// Increment the number of connections for this node by one}'' above a one-line expression with an increment statement. 
\item documentation of a detail: the comment seems relevant, but does not capture the lambda expression as a whole. An example is the comment ``\texttt{// this exception should cause the link chain to explode}'' within a 7-line lambda expression next to a \texttt{throw} statement---while the comment explains this statement, it does not explain the lambda expression as a whole.
\item source code fragment: the comment looks like source code (or pseudocode).
\item other: comments that do not fit into any of the above categories, e.g., TODO comments or comments describing expected output.
\end{itemize}

Another author then used this coding schema to annotate the same 100 comments independently, achieving an agreement of \nicefrac{46}{50} for comments above lambda expressions and \nicefrac{44}{50} for comments within lambda expressions (weighted kappa~\cite{Cohen1960} with five categories: 0.906,\footnote{\url{https://www.graphpad.com/quickcalcs/kappa1/?K=5}, last accessed 22 Jan 2019.} i.e., almost perfect agreement~\cite{Landis1977}). Most of the inconsistent annotations were about ``source code fragment'' vs.~``other'' in cases where the comment \mbox{indicated values}.

\begin{table}
\centering
\caption{Frequency of different kinds of source code comments accompanying lambda expressions}
\label{tab:codecomments}
\begin{tabular}{llr@{\hspace{0.2cm}}r}
\toprule
Location & Type & \multicolumn{2}{c}{Frequency} \\
\midrule
above & high-level documentation & 123 & (62\%)\\
above & reasonably detailed explanation & 22 & (11\%)\\
above & documentation of a detail & 5 & (3\%)\\
above & source code fragment & 11 & (6\%)\\
above & other & 39 & (20\%)\\
\midrule
within & high-level documentation & 0 & (0\%)\\
within & reasonably detailed explanation & 27 & (14\%)\\
within & documentation of a detail & 129 & (65\%)\\
within & source code fragment & 17 & (9\%)\\
within & other & 27 & (14\%)\\
\bottomrule
\end{tabular}
\end{table}

Given the almost perfect agreement, one author then annotated the remaining 150 comments from each group, for a total of 400 annotated source code comments. Table~\ref{tab:codecomments} shows the result of the annotation. The majority of comments above lambda expressions describe the expression, but only at a high level. The majority of comments within lambda expressions document a detail, but cannot be considered as documentation of the entire expression.

This observation encouraged us to employ a rule-based approach for the generation of documentation by \textsc{LambdaDoc}. Given the low quality of existing comments, a machine learning approach appeared infeasible, although future work should confirm this assumption.

\begin{tcolorbox}
\textbf{Summary:} Only 6\% of the lambda expressions in our data set are accompanied by source code comments. Most of these comments describe the lambda expression at a high level or document a detail within the expression.
\end{tcolorbox}

\subsection{\textsc{LambdaDoc} evaluation}
\label{sec:evaluation}

Our manual analysis of the 115 lambda expression documentation attempts produced by the survey participants (5 lambda expressions $\times$ 23 participants) confirmed our conjecture that many developers do not know how to read lambda expressions, which motivated our work on \textsc{LambdaDoc} in the first place. 57 (50\%) of the documentation attempts were inadequate, e.g., ``calculations'' and ``It can be passed around as if it was an object and executed on demand''. Another 39 (34\%) were incomplete, e.g., ``it does compare the variables declared'' and ``This expression takes two parameters and return[s] the result after execution of function''. Only 19 (17\%) of the lambda expression documentation attempts produced by our participants could be considered adequate. Such positive examples include ``compares two integers using splitEvaluation'' (for the second lambda expression in Table~\ref{tab:fivesummaries}) and ``call set with ``c'' for the given ExecutableComponent'' (for the fourth lambda expression in Table~\ref{tab:fivesummaries}).

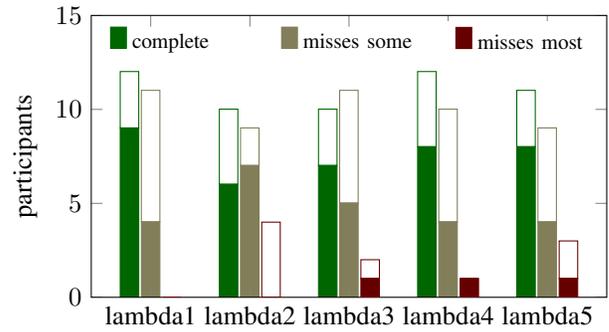
\begin{figure}
\centering
\begin{tikzpicture}[bar width = 7pt]
\begin{axis}[
hide axis,
bar shift = -8pt,
y = 0.25cm,
ymin = 0,
ymax = 15,
ybar stacked,
enlarge x limits = 0.15,
ylabel = participants,
symbolic x coords = {lambda1, lambda2, lambda3, lambda4, lambda5},
xtick = data,
legend style={font=\footnotesize, draw=none, anchor=north west, at={(0.02,0.98)}},
legend entries={complete},
]
\addplot[color=black!60!green, fill=black!60!green] coordinates {(lambda1,9) (lambda2,6) (lambda3,7) (lambda4,8) (lambda5,8)};
\addplot[color=black!60!green] coordinates {(lambda1,3) (lambda2,4) (lambda3,3) (lambda4,4) (lambda5,3)};
\end{axis}
\begin{axis}[
hide axis,
y = 0.25cm,
ymin = 0,
ymax = 15,
ybar stacked,
enlarge x limits = 0.15,
ylabel = participants,
symbolic x coords = {lambda1, lambda2, lambda3, lambda4, lambda5},
xtick = data,
legend style={font=\footnotesize, draw=none, anchor=north, at={(0.5,0.98)}},
legend entries={misses some},
]
\addplot[color=black!60!yellow, fill=black!60!yellow] coordinates {(lambda1,4) (lambda2,7) (lambda3,5) (lambda4,4) (lambda5,4)};
\addplot[color=black!60!yellow] coordinates {(lambda1,7) (lambda2,2) (lambda3,6) (lambda4,6) (lambda5,5)};
\end{axis}
\begin{axis}[
bar shift = 8pt,
y = 0.25cm,
ymin = 0,
ymax = 15,
ybar stacked,
enlarge x limits = 0.15,
ylabel = participants,
symbolic x coords = {lambda1, lambda2, lambda3, lambda4, lambda5},
xtick = data,
legend style={font=\footnotesize, draw=none, anchor=north east, at={(0.98,0.98)}},
legend entries={misses most},
]
\addplot[color=black!60!red, fill=black!60!red] coordinates {(lambda1,0) (lambda2,0) (lambda3,1) (lambda4,1) (lambda5,1)};
\addplot[color=black!60!red] coordinates {(lambda1,0) (lambda2,4) (lambda3,1) (lambda4,0) (lambda5,2)};
\end{axis}
\end{tikzpicture}
\caption{Completeness ratings. Green: The sentence is considered to be complete; Yellow: The sentence misses some important information to understand the lambda expression; Red: The sentence misses the majority of the important information to understand the lambda expression; Solid: Not confident in reading lambda expressions.}
\label{fig:completeness}
\end{figure}

After asking our participants to produce their own documentation, we asked them to assess the documentation generated by \textsc{LambdaDoc} in terms of its completeness, conciseness, and expressiveness (cf.~Table~\ref{tab:surveyquestions}). For our analysis, we distinguished participants based on their self-assessed confidence with regard to reading lambda expressions (cf.~Table~\ref{tab:surveyquestions}, \mbox{Question 7}).

Figure~\ref{fig:completeness} shows the results for completeness. The non-filled part of each bar represents responses by participants who declared themselves to be at least fairly confident in reading lambda expressions. The most positive responses are shown in green, medium responses are shown in yellow, and negative responses are shown in red. As the figure shows, participants generally agreed that the documentation produced by \textsc{LambdaDoc} is complete---the response ``The sentence is considered to be complete'' received the highest number of responses for all but one of the lambda expressions. Responses from participants who considered themselves not confident were slightly more positive---suggesting that \textsc{LambdaDoc} might be especially helpful to newcomers to functional programming in Java.

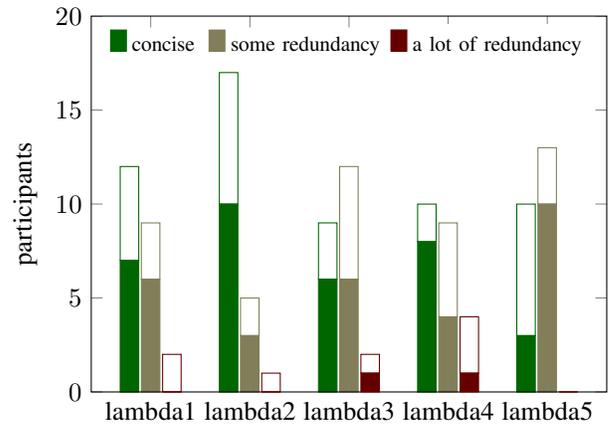
\begin{figure}
\centering
\begin{tikzpicture}[bar width = 7pt]
\begin{axis}[
hide axis,
bar shift = -8pt,
y = 0.25cm,
ymin = 0,
ymax = 20,
ybar stacked,
enlarge x limits = 0.15,
ylabel = participants,
symbolic x coords = {lambda1, lambda2, lambda3, lambda4, lambda5},
xtick = data,
legend style={font=\footnotesize, draw=none, anchor=north west, at={(0.02,0.98)}},
legend entries={concise},
]
\addplot[color=black!60!green, fill=black!60!green] coordinates {(lambda1,7) (lambda2,10) (lambda3,6) (lambda4,8) (lambda5,3)};
\addplot[color=black!60!green] coordinates {(lambda1,5) (lambda2,7) (lambda3,3) (lambda4,2) (lambda5,7)};
\end{axis}
\begin{axis}[
hide axis,
y = 0.25cm,
ymin = 0,
ymax = 20,
ybar stacked,
enlarge x limits = 0.15,
ylabel = participants,
symbolic x coords = {lambda1, lambda2, lambda3, lambda4, lambda5},
xtick = data,
legend style={font=\footnotesize, draw=none, anchor=north, at={(0.4,0.98)}},
legend entries={some redundancy},
]
\addplot[color=black!60!yellow, fill=black!60!yellow] coordinates {(lambda1,6) (lambda2,3) (lambda3,6) (lambda4,4) (lambda5,10)};
\addplot[color=black!60!yellow] coordinates {(lambda1,3) (lambda2,2) (lambda3,6) (lambda4,5) (lambda5,3)};
\end{axis}
\begin{axis}[
bar shift = 8pt,
y = 0.25cm,
ymin = 0,
ymax = 20,
ybar stacked,
enlarge x limits = 0.15,
ylabel = participants,
symbolic x coords = {lambda1, lambda2, lambda3, lambda4, lambda5},
xtick = data,
legend style={font=\footnotesize, draw=none, anchor=north east, at={(0.98,0.98)}},
legend entries={a lot of redundancy},
]
\addplot[color=black!60!red, fill=black!60!red] coordinates {(lambda1,0) (lambda2,0) (lambda3,1) (lambda4,1) (lambda5,0)};
\addplot[color=black!60!red] coordinates {(lambda1,2) (lambda2,1) (lambda3,1) (lambda4,3) (lambda5,0)};
\end{axis}
\end{tikzpicture}
\caption{Conciseness ratings. Green: The sentence is considered to be concise; Yellow: The sentence contains some redundant/useless information; Red: The sentence contains a lot of redundant/useless information; Solid: Not confident in reading lambda expressions.}
\label{fig:conciseness}
\end{figure}

Figure~\ref{fig:conciseness} shows the participant responses with regard to conciseness. For most of the lambda expressions, the positive answer ``The sentence is considered to be concise'' was selected most often, while the documentation generated for the third and fifth expression was considered to contain some \mbox{redundant/useless} information. As the documentation in Table~\ref{tab:fivesummaries} shows, we opted for \textsc{LambdaDoc} to produce precise and detailed documentation---future work should explore whether there exists a better balance between precision and conciseness. Very few participants indicated that \textsc{LambdaDoc} produces a lot of redundant/useless information.

\begin{figure}
\centering
\begin{tikzpicture}[bar width = 7pt]
\begin{axis}[
hide axis,
bar shift = -8pt,
y = 0.25cm,
ymin = 0,
ymax = 17,
ybar stacked,
enlarge x limits = 0.15,
ylabel = participants,
symbolic x coords = {lambda1, lambda2, lambda3, lambda4, lambda5},
xtick = data,
legend style={font=\footnotesize, draw=none, anchor=north west, at={(0.02,0.98)}},
legend entries={easy to read},
]
\addplot[color=black!60!green, fill=black!60!green] coordinates {(lambda1,7) (lambda2,8) (lambda3,7) (lambda4,9) (lambda5,6)};
\addplot[color=black!60!green] coordinates {(lambda1,6) (lambda2,3) (lambda3,4) (lambda4,5) (lambda5,5)};
\end{axis}
\begin{axis}[
hide axis,
y = 0.25cm,
ymin = 0,
ymax = 17,
ybar stacked,
enlarge x limits = 0.15,
ylabel = participants,
symbolic x coords = {lambda1, lambda2, lambda3, lambda4, lambda5},
xtick = data,
legend style={font=\footnotesize, draw=none, anchor=north, at={(0.5,0.98)}},
legend entries={somewhat readable},
]
\addplot[color=black!60!yellow, fill=black!60!yellow] coordinates {(lambda1,5) (lambda2,5) (lambda3,5) (lambda4,3) (lambda5,7)};
\addplot[color=black!60!yellow] coordinates {(lambda1,3) (lambda2,6) (lambda3,3) (lambda4,3) (lambda5,2)};
\end{axis}
\begin{axis}[
bar shift = 8pt,
y = 0.25cm,
ymin = 0,
ymax = 17,
ybar stacked,
enlarge x limits = 0.15,
ylabel = participants,
symbolic x coords = {lambda1, lambda2, lambda3, lambda4, lambda5},
xtick = data,
legend style={font=\footnotesize, draw=none, anchor=north east, at={(0.98,0.98)}},
legend entries={hard to read},
]
\addplot[color=black!60!red, fill=black!60!red] coordinates {(lambda1,1) (lambda2,0) (lambda3,1) (lambda4,1) (lambda5,0)};
\addplot[color=black!60!red] coordinates {(lambda1,1) (lambda2,1) (lambda3,3) (lambda4,2) (lambda5,3)};
\end{axis}
\end{tikzpicture}
\caption{Expressiveness ratings. Green: The sentence is easy to read and understand; Yellow: The sentence is somewhat readable and understandable; Red: The sentence is hard to read and understand; Solid: Not confident in reading lambda expressions.}
\label{fig:expressiveness}
\end{figure}
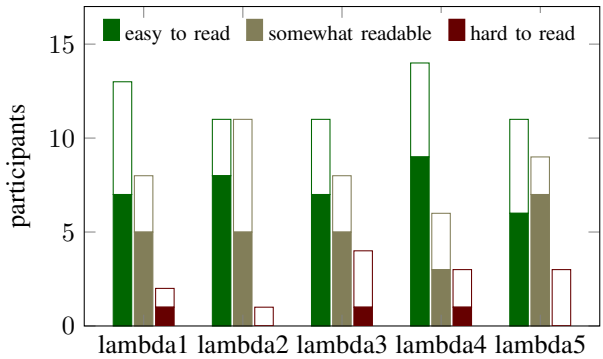

The answer ``The sentence is easy to read and understand'' was selected most often in response to our question about the expressiveness of the content generated by \textsc{LambdaDoc}, as shown in Figure~\ref{fig:expressiveness}. Only our treatment of array indices which affected the documentation of the second lambda expression (cf.~Table~\ref{tab:fivesummaries}) prompted an equal number of ``somewhat readable and understandable'' ratings. As with the previous questions, the differences between participants confident in reading lambda expressions and those who indicated not be confident appear negligible.

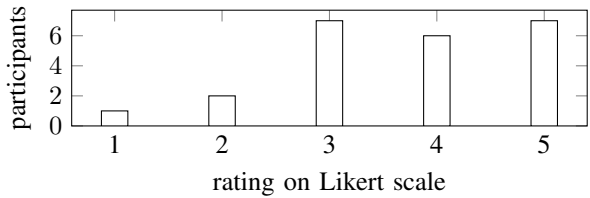
\begin{figure}
\centering
\begin{tikzpicture}
\begin{axis}[
    y = 0.2cm,
    ylabel = participants,
    xlabel = rating on Likert scale,
    symbolic x coords={1, 2, 3, 4, 5},
    xtick=data,
    ymin = 0]
    \addplot[ybar,fill=white] coordinates {
    (1, 1)
    (2, 2)
    (3, 7)
    (4, 6)
    (5, 7)
    };
\end{axis}
\end{tikzpicture}
\caption{Survey responses to ``How likely is it that you would recommend this tool to a friend or colleague?''. 1 -- ``not likely at all''; 5 -- ``extremely likely''.}
\label{fig:recommendation}
\end{figure}

Participants also left positive feedback after trying the online version of our tool, e.g., ``Very helpful in adapting to a new way of writing code''. Figure~\ref{fig:recommendation} shows the participants' responses to our survey question ``How likely is it that you would recommend this tool to a friend or colleague?'' on a 5-point Likert scale. The majority of participants were positive about issuing such a recommendation. Finally, responses as to how participants would like to have \textsc{LambdaDoc} implemented varied from GUI application (11) and website / web service (10) to Eclipse plugin (9, multiple answers possible). Similarly, participants could see themselves using \textsc{LambdaDoc} for documentation (13), testing (10), implementation (8), and maintenance (4).

\begin{tcolorbox}
\textbf{Summary:} When asked to document lambda expressions, most of the responses produced by our participants were inadequate. In contrast, the documentation produced by \textsc{LambdaDoc} was largely perceived to be complete, concise, and easy to read and understand.
\end{tcolorbox}

\section{Threats to Validity}
\label{sec:threats}

As with all empirical studies, there are a number of threats that may affect the validity of our results. 

Threats to construct validity concern the suitability of our evaluation metrics. Following the work of Linares-V\'{a}squez et al.~\cite{Linares-Vasquez2016} and others, we evaluated the documentation generated by \textsc{LambdaDoc} in terms of its perceived completeness, conciseness, and expressiveness, using survey questions very similar to those used in previous work. Future work should explore other dimensions of the generated documentation, e.g., helpfulness. Our algorithm for detecting lambda expressions could potentially lead to inaccuracies. However, the findings from our first research question show that in a statistically representative sample, all lambda expressions were correctly identified. It is also possible that another group of researchers would have identified different kinds of documentation in the source code comments which accompanies lambda expressions. However, our inter-rater agreement was almost perfect, increasing the confidence in our findings.

Threats to external validity affect the generalisability of our findings. We cannot claim that our findings generalise beyond the particular data set we have considered in this work. Our work may not generalise to other programming languages or other functional programming constructs. The number of study participants and the number of lambda expressions used in the evaluation of \textsc{LambdaDoc} are also necessarily limited. Asking different participants about their perceptions of \textsc{LambdaDoc} might have resulted in different findings. All lambda expressions used in the evaluation (cf.~Table~\ref{tab:fivesummaries}) were single-line expressions. While we found that single-line lambda expressions are in the majority on GitHub (cf.~Figure~\ref{fig:linesperlambda}), future work should investigate the perceived completeness, conciseness, and expressiveness of documentation generated for multi-line lambda expressions.

Threats to internal validity relate to errors or inaccuracies in our implementation. Our current implementation is unable to detect embedded lambda expressions if both arrows are on the same line. Apart from this issue, we have double-checked our source code and fixed all errors we found. Still, there could be additional errors which we did not notice.

\section{Related Work}
\label{sec:related}

After the introduction of lambda expressions, prior work started to investigate the use of lambda expressions and its impact.
Uesbeck et al.~found that using lambda expressions in C++ has a negative impact on programming speed of inexperienced users~\cite{Uesbeck2016}.
To encourage developers to adopt new language features, Khatchadourian and Masuhara submitted pull requests introducing language features to open source projects~\cite{Khatchadourian2018}.
Mazinanian et al.~investigated the adoption of lambda expressions in 241 Java open source projects and found that projects migrate to lambda expressions by converting classes to lambda expressions, replacing loops/conditionals with streams, and enhancing functionality by wrapping existing code to lambda expressions~\cite{Mazinanian2017}.
Complementing the prior work, in this work, we found that exception handling is a common purpose of using lambda expressions in Java and that lambda expressions are usually implicit, single-line, and have one parameter.

While software documentation makes it easier for developers to comprehend software artefacts, manually-written documentation becomes a tedious task for developers.
Several studies empirically investigate the essentials of software documentation.
De Souza et al.~showed that  developers perceived that source code comments are the second most important software artefact in Agile software development\cite{deSouza2005}.
Lin et al.~found that developers spent effort on maintaining API documentation, e.g., literal polishes~\cite{Lin2011}.
Fluri et al.~found that newly added code was rarely documented and a source code comment was often changed along with the associated source code~\cite{Fluri2007}.
Li et al.~reported that while developers need documentation to understand unit test cases, a large proportion of C\# projects on GitHub lacked comments for unit test cases~\cite{Li2016}.
Moreover, Ibrahim et al.~found that neglecting to update a comment increases the probability of having future defects in a software system~\cite{Ibrahim2012}.
In this paper, we empirically investigated the source code comments accompanying lambda expressions and found that only 6\% of the lambda expressions have corresponding source code comments.


Several researchers have developed approaches to automatically summarise Java source code.
For example, 
Moreno et al.~developed an approach to summarise the information and details of Java classes~\cite{Moreno2013}.
McBurney et al.~analysed the method calls and leveraged the PageRank algorithm to generate a description of the behaviour of a Java method~\cite{McBurney2016}.
Ying and Robillard developed an approach for the automated summarisation of code fragments~\cite{Ying2013}.
Buse et al.~developed an approach to automatically summarise conditions of Java exceptions~\cite{Buse2008}.
Furthermore, automated documentation generation has been developed for other software artefacts.
Prior work developed an approach to automatically summarise unit test cases~\cite{Li2016} and test failures~\cite{Sai2011}. 
To help developers understand database schemata when writing database-related code, Linares-V\'{a}squez et al.~developed an approach to analyse database schemata and SQL statements to automatically describe database usage at the source code method level~\cite{Linares-Vasquez2016}.
Racchetti et al.~proposed an approach to automatically generate documentation for Programmable Logic Controller (PLC) code~\cite{Racchetti2015}.
Hassan and Hill presented a technique towards automatically generating comments for Java statements suitable for novice programmers~\cite{Hassan2018}.
Moreover, prior studies proposed an approach to summarise individual code changes~\cite{Buse2010} and the software evolution based on code changes~\cite{Kim2013}.
Recently, Robillard et al.~outlined a research agenda for generating developer documentation on-demand~\cite{Robillard2017}.
In this work, we are the first to develop an approach to automatically generate documentation for lambda expressions in Java.


\section{Conclusions and Future Work}
\label{sec:conclusions}

The lines between functional programming languages and object-oriented programming languages have become blurred with the addition of functional programming constructs to non-functional programming languages. One example is the introduction of lambda expressions to Java as part of the release of Java 8 in 2014. If used correctly, lambda expressions can reduce the amount of code required for certain tasks, and they allow for greater efficiency through sequential and parallel execution support. However, since lambda expressions are still a relatively new language feature, not all developers use or understand them.

To help developers read lambda expressions and understand how they are used, we have presented \textsc{LambdaDoc} which automatically detects lambda expressions in a Java repository and can generate natural language documentation for them. We evaluated \textsc{LambdaDoc} with 23 professional developers and found that the generated documentation was perceived to be complete, concise, and expressive. In contrast, when asked to document lambda expressions, most of the responses produced by our participants were inadequate.

We augmented the introduction and evaluation of \textsc{LambdaDoc} with the results of an empirical study which found that 11\% of Java GitHub repositories make use of lambda expressions and that exception handling is a common purpose of using the construct. We also found that only 6\% of the lambda expressions in our data set were accompanied by a source code comment.

In our future work, we will extend our empirical study to take the evolution of lambda expression use into account. We plan to deploy \textsc{LambdaDoc} as a web service which integrates with GitHub, to make it available to more developers and further evaluate the tool. We will also design improvements to the documentation generated by \textsc{LambdaDoc}, to make the documentation more concise and to resolve types using call graph analysis. In addition, we will explore the integration of \textsc{LambdaDoc} with source code summarisation approaches (e.g., McBurney and McMillan~\cite{McBurney2016}) to generate concise documentation for multi-line lambda expressions. We hope that these efforts will further increase the adoption of functional programming constructs in Java, enabling developers to take advantage of the potential performance and readability improvements which originally motivated Java's foray into the realm of functional programming.

\section*{Acknowledgements}

We thank all participants for their feedback on \textsc{LambdaDoc}. We thank the Government of Saudi Arabia for supporting the first author's studies. This work has been supported by the Australian Research Council's Discovery Early Career Researcher Award (DECRA) funding scheme (DE180100153). 

\bibliographystyle{IEEEtran}
\IEEEtriggeratref{15}
\bibliography{main}

\end{sloppy}
\end{document}